\documentclass[a4paper]{article}

\usepackage{a4wide}
\usepackage[UKenglish]{babel}
\usepackage{graphicx,subfigure,epstopdf}
	\graphicspath{{./}{figure/}}
\usepackage[colorlinks=true,linkcolor=blue,citecolor=red]{hyperref}
\usepackage{xcolor}
\usepackage{amsmath,amsthm,bbold}
\usepackage{algorithm,algorithmic}
\usepackage{enumitem}
\usepackage{array}

\newcommand{\abs}[1]{\left\lvert#1\right\rvert}
\newcommand{\ave}[1]{\left\langle#1\right\rangle}
\newcommand{\N}{\mathbb{N}}
\newcommand{\bn}{\mathbf{n}}
\renewcommand{\P}{\operatorname{Prob}}

\newcommand{\R}{\mathbb{R}}
\renewcommand{\S}{\mathbb{S}}

\newcommand{\bv}{\mathbf{v}}
\newcommand{\bV}{\mathbf{V}}
\newcommand{\sV}{\mathcal{V}}
\newcommand{\VV}{\mathbb{V}}

\allowdisplaybreaks

\begin{document}
\title{The Boltzmann legacy revisited: kinetic models of social interactions}
\author{Martina Fraia\thanks{\texttt{s237819@studenti.polito.it}} \and Andrea Tosin\thanks{\texttt{andrea.tosin@polito.it}}}
\date{\small Department of Mathematical Sciences ``G. L. Lagrange'' \\ Politecnico di Torino, Italy}

\maketitle

\begin{abstract}
The application of classical methods of statistical mechanics, originally developed by Ludwig Boltzmann in gas dynamics, to the description of social phenomena is a successful story that we try to outline in this paper. On one hand, it is nowadays a flourishing research line, which is more and more permeating different contexts such as the econophysics, sociophysics, biomathematics, transportation engineering to name just a few of them. On the other hand, it is a fascinating mathematical challenge, because it requires the interplay of various complementary expertises: modelling, model analysis, numerics. In this paper, we try to give a taste of all of this using the social phenomenon of opinion formation as a motivating example.

\medskip

\begin{center}
\textbf{Sommario}
\end{center}
L'applicazione dei metodi classici della meccanica statistica, sviluppati originariamente da Ludwig Boltzmann per la gasdinamica, alla descrizione di fenomeni sociali \`{e} una storia di successo che in questo articolo cerchiamo di tratteggiare. Da un lato essa costituisce attual\-mente una fiorente linea di ricerca, che sta sempre pi\`{u} permeando contesti diversi tra loro quali l'econofisica, la sociofisica, la biomatematica, l'ingegneria dei trasporti per non citare che alcuni esempi. Dall'altro \`{e} anche una sfida matematica affascinante, perch\'{e} richiede l'interazione di svariate competenze complementari: la modellistica, l'analisi dei modelli, la numerica. In questo articolo cerchiamo di dare un assaggio di tutto ci\`{o} usando come esempio motivante la formazione delle opinioni.

\bigskip


\medskip

\end{abstract}

\section{Introduction}
\begin{figure}[!t]
\centering
\includegraphics[scale=0.8]{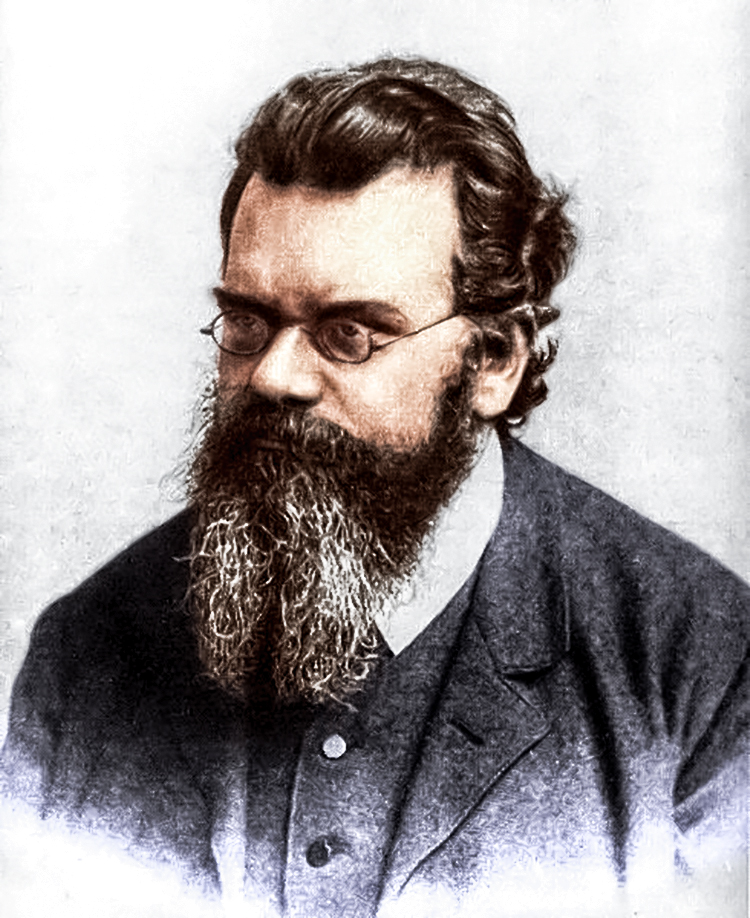}
\caption{Ludwig E. Boltzmann was born in Vienna in $1844$. He proposed to use probability to describe how the collisions among the atoms determine the aggregate properties of the matter. His work marked the beginning of \textit{statistical mechanics}, although at his time many scientists did not trust his ideas. He died in Duino, near Trieste, in $1906$. (Picture source: Wikimedia Commons.)}
\label{fig:boltzmann}
\end{figure}

In the late 1800 the Austrian physicist Ludwig Eduard Boltzmann (Figure~\ref{fig:boltzmann}) formulated the celebrated equation bearing his name to explain the complex concepts of thermodynamics starting from the simple mechanics of colliding gas molecules. Nowadays the legacy of his theory has imposed itself in contexts very distant from the original one, such as e.g., socio-economic dynamics, however with an analogous goal: to understand and explain how the collective behaviour of human societies originates from simple individual interactions empirically experienceable in personal lives.

On one hand, the contemporary research in kinetic theory still focuses on the study of fundamental mathematical-physical properties of the Boltzmann equation, see e.g.,~\cite{bobylev2018KRM,james2019JNL,pulvirenti2017IM,villani1998ARMA}, which from the theoretical point of view are not always fully understood yet. Nevertheless these works are quite technical, hence they might be hardly accessible to a non-specialised readership. For an introduction to the mathematical topics of the Boltzmann equation, we recommend instead the lecture notes~\cite{bressan2005LN,villani2002HMFD} and the book~\cite{cercignani2000BOOK}.

On the other hand, starting approximately from the early $2000$, other research paths have been opened, which focus instead on the application of the Boltzmann paradigm to systems of interacting ``particles'' different from gas molecules and parallelly deal with the new physical and mathematical issues that these applications raise~\cite{pareschi2013BOOK}. It is the case of the socio-economic applications mentioned above, which provide a breeding ground for a fruitful and innovative revisitation of classical concepts and methods of kinetic theory. At the same time, the kinetic theory offers a sound scientific background to formalise quantitatively the descriptions of those systems, which are often heuristic, qualitative and sometimes also biased by personal ideological views.

In this paper we aim to provide an overview of the way in which some classical concepts of the kinetic theory for gas dynamics have been evolved into mathematical tools suitable to model the so-called \textit{multi-agent systems} and understand the fundamental links between their individual and collective behaviour. In particular, in Section~\ref{sect:legacy} we quickly review the physical and mathematical ideas behind the original Boltzmann equation for colliding gas molecules. In Section~\ref{sect:interactions} we discuss how a parallelism between those classical ideas and entirely different interactions may be established, considering as a prototype the exchange of opinions among people. In Section~\ref{sect:sznajd_revisited} we exemplify the application of the kinetic methods to a simple model of opinion formation, which we obtain by revisiting the classical Sznajd model~\cite{sznajd-weron2000IJMP}. The latter is particularly inspiring, because it springs from another classical model, the Ising model~\cite{ising1925ZFP}, which was conceived to describe the magnetism of the matter. Therefore it fits particularly well into our discussion, because it constitutes in turn a limpid example of how, in modern Applied Mathematics, classical phenomena may serve as the basis on which to ground the description of non-classical ones, which typically lack the necessary background theories. The qualitative results of this opinion formation model are the occasion to discuss, in Section~\ref{sect:interlude}, to which extent the main goal of a mathematical model is the accurate reproduction of empirical observations; and to stress what much greater informative value even a simple toy model may instead have if it is built on sound mathematical bases. Along this line, in Section~\ref{sect:indecisiveness} we propose a simple yet natural generalisation of the opinion formation model presented in Section~\ref{sect:sznajd_revisited}, which proves the flexibility of the Boltzmann-type kinetic approach in describing social phenomena. Finally, we collect in Appendix~\ref{sect:derivation} a slightly technical but instructive derivation of the Boltzmann-type equation, which may provide to the interested readers further insights into the physical and mathematical meaning of the kinetic description of multi-agent systems.

\section{The Boltzmann legacy}
\label{sect:legacy}
In the Boltzmann kinetic theory of gases, the microscopic state of a gas molecule is represented by the velocity $\bv\in\R^3$, which may change in consequence of collisions with other molecules. Such collisions are assumed to be elastic, therefore they are described by appealing to the classical conservations of momentum and kinetic energy. A further distinguishing assumption is that collisions are mainly \textit{binary}, i.e. they involve only two molecules at a time. Collisions among three or more molecules are disregarded as higher order effects, i.e. much less probable events than binary collisions. Hence, if $\bv,\,\bv_\ast\in\R^3$ are the \textit{pre-collisional} velocities of two colliding molecules and $\bv',\,\bv_\ast'\in\R^3$ their \textit{post-collisional} velocities instantaneously produced by an elastic collision, the following laws hold true:
$$
	\begin{cases}
		m\bv+m\bv_\ast=m\bv'+m\bv_\ast' \\
		\frac{1}{2}m\abs{\bv}^2+\frac{1}{2}m\abs{\bv_\ast}^2=\frac{1}{2}m\abs{\bv'}^2+\frac{1}{2}m\abs{\bv_\ast'}^2,
	\end{cases}
$$
where, for simplicity, the gas molecules are assumed to have all the same mass $m>0$. These relationships allow one to express the post-collisional velocities in terms of the pre-collisional velocities as
\begin{equation}
	\begin{cases}
		\bv'=\bv+((\bv_\ast-\bv)\cdot\bn)\bn \\
		\bv_\ast'=\bv_\ast+((\bv-\bv_\ast)\cdot\bn)\bn,
	\end{cases}
	\label{eq:collisions}
\end{equation}
$\bn$ being a unit vector such that $(\bv-\bv_\ast)\cdot\bn\geq 0$. We will write this as $\bn\in\S^2_+$.

Equations~\eqref{eq:collisions} are valid for a generic pair of colliding molecules of the gas. In other words, the collision described by~\eqref{eq:collisions} is representative of any possible collision between two gas molecules, because the latter are assumed to be \textit{indistinguishable}. As such,~\eqref{eq:collisions} is the basis for a \textit{statistical} description of the superposition of numerous collisions of that type, which is obtained by regarding the velocity of a generic molecule as a random variable $\bV_t\in\R^3$, where $t>0$ is the time, distributed according to a probability density function $f=f(t,\,\bv)$ such that
$$ \P(\bV_t\in A)=\int_Af(t,\,\bv)\,d\bv $$
for all (measurable) set $A\subseteq\R^3$. The evolution in time of $\bV_t$ is essentially ruled by~\eqref{eq:collisions} and entails a corresponding evolution in time of $f$. The celebrated \textit{Boltzmann equation} is precisely the mathematical determination of the evolution of $f$ under~\eqref{eq:collisions}. It may be written as (cf. e.g.,~\cite[Chapt.~1]{pareschi2013BOOK})
\begin{align}
	\begin{aligned}[b]
		\partial_tf(t,\,\bv) &= Q(f,\,f)(t,\,\bv) \\
		&:= \int_{\R^3}\int_{\S^2_+}\abs{(\bv-\bv_\ast)\cdot\bn}\bigl(f(t,\,\bv')f(t,\,\bv_\ast')-f(t,\,\bv)f(t,\,\bv_\ast)\bigr)\,d\bn\,d\bv_\ast,
	\end{aligned}
	\label{eq:boltzmann}
\end{align}
where the operator $Q$ on the right-hand side, called the \textit{collision operator}, expresses the average effect of many interactions of the form~\eqref{eq:collisions} on the time variation of the velocity distribution $f$. The term $\abs{(\bv-\bv_\ast)\cdot\bn}$ is the \textit{collision kernel}: it models the rate at which any two molecules collide depending on their relative pre-collisional velocity $\bv-\bv_\ast$. Since $Q(f,\,f)$ is an integral operator, the Boltzmann equation~\eqref{eq:boltzmann} turns out to be an \textit{integro-differential equation}.

The collision operator $Q$ features the quadratic non-linearities $f(t,\,\bv')f(t,\,\bv_\ast')$, $f(t,\,\bv)f(t,\,\bv_\ast)$. They are the result of a further assumption which strongly characterises the Boltzmann approach, that of \textit{molecular chaos}. In principle, when computing the average effect of the collisions~\eqref{eq:collisions} one should use the joint probability density $f_2=f_2(t,\,\bv,\,\bv_\ast)$ of the velocities of the colliding molecules. Nevertheless, in this way the time evolution of $f$ would depend on the further unknown $f_2$. However, considering that the past collisional story of any two colliding molecules will consist likely of many other collisions with a large number of different molecules, one may assume that the two colliding molecules are statistically independent at the moment of the collision and hence
$f_2(t,\,\bv,\,\bv_\ast)=f(t,\,\bv)f(t,\,\bv_\ast)$. This assumption, also known as \textit{Boltzmann Ansatz}, is particularly justified in the case of a gas, in view of the extremely high number of molecules composing it.

\section{From molecule collisions to social interactions}
\label{sect:interactions}
Starting approximately from the early $2000$, the ideas of statistical mechanics at the basis of the Boltzmann kinetic approach to gas dynamics have been intensely revisited and applied to systems of interacting particles possibly quite different from a gas. In particular, one of the most fascinating applications has been to various types of human behaviour.

Probably one of the first contributions in this direction was the one by the Russian physicist Ilya Prigogine, who already in the early sixties proposed a Boltzmann-type kinetic description of road traffic, in particular of the interactions among vehicles in a traffic stream~\cite{prigogine1961PROC,prigogine1960OR,prigogine1971BOOK}. Such interactions are still essentially mechanical, because they are formalised in terms of accelerations and decelerations of a vehicle depending on the speed of the leading vehicle. Nevertheless, at the same time they are different from~\eqref{eq:collisions}, because they are not collisions in the classical physical sense. Indeed, vehicles need not collide to change their speed, rather they change speed in order not to collide. Even more, unlike~\eqref{eq:collisions} these interactions are essentially \textit{heuristic}, because for vehicular traffic there is not a background physical theory providing a fundamental model of car (viz. driver) behaviour. The same difficulty had to be faced in the case of other types of social interactions, i.e. interactions relying essentially on human behaviour. We mention, for instance, trading-type interactions leading to the redistribution of wealth in a society and the emergence of income distribution curves~\cite{cordier2009JSP,cordier2005JSP,slanina2004PRE} or interactions producing the formation and spreading of opinions~\cite{ochrombel2001IJMP,sznajd-weron2000IJMP,toscani2006CMS}. In the latter case, an additional difficulty is that the \textit{opinion} of an individual is not a well-defined and measurable physical quantity.

Let $v$ be the \textit{social state} of a representative particle in a population of interacting individuals. With reference to the examples above, $v$ may be the speed, the wealth or the opinion. In general, $v$ is assumed to be a scalar variable belonging to a certain subset $\sV$ of the real axis. When $v$ represents an opinion, like in the applications that we will discuss in the forthcoming Sections~\ref{sect:sznajd_revisited},~\ref{sect:indecisiveness}, it is customary to take $\sV=[-1,\,1]$ and to understand the states $v=\pm 1$ as two opposite extreme convictions and $v=0$ as the indecisiveness. Inspired by the collisions~\eqref{eq:collisions}, we may describe an interaction between two individuals with social states $v,\,v_\ast\in\sV$ as an update rule of the form:
\begin{equation}
	\begin{cases}
		v'=v+I(v,\,v_\ast) \\
		v_\ast'=v_\ast+I_\ast(v_\ast,\,v),
	\end{cases}
	\label{eq:interactions}
\end{equation}
where $I,\,I_\ast$ are two possibly different interaction functions defined on $\sV\times\sV$. If $I=I_\ast$ then the interactions~\eqref{eq:interactions} are said to be \textit{symmetric}, because the second one is obtained from the first one by simply switching $v$ and $v_\ast$. In order for~\eqref{eq:interactions} to be physically consistent it is necessary that $I$, $I_\ast$ are chosen in such a way to guarantee $v',\,v_\ast'\in\sV$ for all $v,\,v_\ast\in\sV$. We observe that this issue is not present in model~\eqref{eq:collisions}, because there the microscopic states of the molecules belong to the whole space $\R^3$. Conversely, in~\eqref{eq:interactions} this issue arises whenever $\sV$ is a proper subset of $\R$, thereby making a first relevant technical difference with respect to the classical framework recalled in Section~\ref{sect:legacy}.

For instance, in~\cite{toscani2006CMS} a model for \textit{opinion consensus} is proposed in the form~\eqref{eq:interactions} with
\begin{equation}
	I(v,\,v_\ast)=\gamma(v_\ast-v), \qquad I_\ast(v_\ast,\,v)=\gamma(v-v_\ast),
	\label{eq:I}
\end{equation}
where $\gamma>0$ is a parameter. In general, the functions~\eqref{eq:I} do not ensure that $v',\,v_\ast'\in [-1,\,1]$ for all $v,\,v_\ast\in [-1,\,1]$. Assume indeed that two individuals with pre-interaction opinions $v=0$, $v_\ast=1$ meet. Then from~\eqref{eq:interactions} with~\eqref{eq:I} we compute $v'=\gamma$ and $v_\ast'=1-\gamma$, whence we see that if $\gamma>1$ then $v',\,v_\ast'\not\in [-1,\,1]$. On the other hand, it is not difficult to prove that $\gamma\leq 1$ is a necessary and sufficient condition to guarantee the physical consistency of all the interactions.

Exactly like in the case of the gas molecules,~\eqref{eq:interactions} is the basis for an aggregate description of the system of interacting individuals. If we had to simulate the statistical evolution of the system according to~\eqref{eq:interactions}, we could proceed conceptually as follows:
\begin{algorithm}[!h]
	\caption{}
	\begin{algorithmic}[1]
		\STATE Assume that, at some time $n\in\N$, we have a sufficiently large sample of opinions:
			$$ \VV^n=\left\{v_1^n,\,v_2^n,\,\dots,\,v_N^n\right\}, $$
			$N\gg 1$ being the size of the sample
		\FOR{$n=0,\,1,\,2,\,\dots$}
			\REPEAT
			\STATE Pick randomly (e.g., uniformly) two different opinions $v_i^n,\,v_j^n\in\VV^n$
			\STATE \label{state:decide} Decide if the agents $i$, $j$ interact, for instance by tossing a (possibly biased) coin
			\IF{the agents $i$, $j$ interact}
				\STATE Update $v_i^n,\,v_j^n$ according to~\eqref{eq:interactions}:
				\begin{equation*}
					\begin{cases}
						v_i^{n+1}=v_i^n+I(v_i^n,\,v_j^n) \\
						v_j^{n+1}=v_j^n+I_\ast(v_j^n,\,v_i^n)
					\end{cases}
				\end{equation*}
			\ELSE
				\STATE Leave the opinions unchanged:
				\begin{equation*}
					\begin{cases}
						v_i^{n+1}=v_i^n \\
						v_j^{n+1}=v_j^n
					\end{cases}
				\end{equation*}
			\ENDIF
			\STATE Discard from $\VV^n$ the pair of opinions just used
			\UNTIL{no unused opinions are left in $\VV^n$}
			\STATE Form the new sample $\VV^{n+1}:=\left\{v_1^{n+1},\,v_2^{n+1},\,\dots,\,v_N^{n+1}\right\}$
			\STATE \label{state:hist} Build a histogram from the data in $\VV^{n+1}$, which depicts the new statistical distribution of the opinions
		\ENDFOR
	\end{algorithmic}
	\label{alg:nanbu}
\end{algorithm}

In order to start this iterative procedure, we need to generate a first sample of opinions $\VV^0=\{v_1^0,\,v_2^0,\,\dots,\,v_N^0\}$ representing the \textit{initial condition} of the system. This may be done, for instance, by sampling uniformly $N$ random numbers in $[-1,\,1]$ with the appropriate built-in routines available in most programming languages. As a matter of fact, this corresponds to initialise the system with the opinions uniformly distributed according to the probability density
$$ f^0(v)=\frac{1}{2}\chi(-1\leq v\leq 1), $$
where
$$ \chi(v\in A)=
	\begin{cases}
		1 & \text{if } v\in A \\
		0 & \text{otherwise}
	\end{cases}
$$
is the characteristic function of the set $A\subseteq\R$. To prescribe a different initial opinion distribution, it is necessary to build $\VV^0$ by sampling from a non-uniform probability density function. For this, one may take advantage of suitable \textit{ad hoc} numerical methods, cf. e.g.,~\cite{pareschi2005ESAIMP}.

The histogram built in step~\ref{state:hist} of Algorithm~\ref{alg:nanbu} is an approximation of the probability density function $f^n(v)$ of the opinions after $n$ iterations of the dynamics~\eqref{eq:interactions}. It is only an approximation because it is computed from a sequence of particular realisations of the interactions~\eqref{eq:interactions}, starting from a particular sample of the initial distribution $f^0(v)$ among all possible ones. In general, it is therefore important to obtain also a \textit{mathematical model} of the evolution of the true opinion distribution $f(t,\,v)$ under the interaction rules~\eqref{eq:interactions}. In~\cite[Sects. 1.3.1,~1.3.2,~1.4.1]{pareschi2013BOOK} the authors illustrate a constructive procedure to obtain formally an equation for $f$. The procedure is illuminating, because it unveils the links between the simple physics of random individual interactions and their aggregate statistical description, but is slightly technical. Therefore we defer it to Appendix~\ref{sect:derivation} for the interested readers. Here we report instead the final equation, which reads:
\begin{equation}
	\frac{d}{dt}\int_{\sV}\varphi(v)f(t,\,v)\,dv
		=\frac{1}{2}\int_{\sV}\int_{\sV}B\bigl(\varphi(v')+\varphi(v_\ast')-\varphi(v)-\varphi(v_\ast)\bigr)f(t,\,v)f(t,\,v_\ast)\,dv\,dv_\ast
	\label{eq:boltzmann-type}
\end{equation}
and is called a \textit{Boltzmann-type equation}. At first glance, the analogy with~\eqref{eq:boltzmann} is actually not evident, except for the quadratic non-linearity $f(t,\,v)f(t,\,v_\ast)$ on the right-hand side. First of all, we need to specify that the function $\varphi$ appearing in~\eqref{eq:boltzmann-type} is a test function. This means that~\eqref{eq:boltzmann-type} is technically a weak form of the equation for $f$, i.e. an equation which is required to hold for every possible choice of $\varphi$. In Appendix~\ref{sect:derivation} we show that, upon passing from~\eqref{eq:boltzmann-type} to the corresponding strong form, the analogy with the classical Boltzmann equation~\eqref{eq:boltzmann} becomes much more apparent. Here we want to focus instead on the fact that~\eqref{eq:boltzmann-type} has an interesting and instructive physical interpretation, which explains clearly the intuitive idea translated by~\eqref{eq:boltzmann-type} despite its apparently complicated form. Let us think of $\varphi$ as an \textit{observable quantity}, i.e. any quantity that can be computed out of the knowledge of the opinion $v$ of an agent. We may recognise that $\int_{\sV}\varphi(v)f(t,\,v)\,dv$ on the left-hand side is the mean of the observable quantity $\varphi$ at time $t$. For example, if we choose $\varphi(v)=v^m$ for some integer $m\in\N$ then $\int_{\sV}\varphi(v)f(t,\,v)\,dv$ becomes the $m$-th statistical moment of the distribution of the opinions. On the other hand, $\frac{1}{2}\left(\varphi(v')+\varphi(v_\ast')-\varphi(v)-\varphi(v_\ast)\right)$ on the right-hand side is the mean variation of $\varphi$ in a single interaction~\eqref{eq:interactions}. Therefore, the concept expressed by the Boltzmann-type equation~\eqref{eq:boltzmann-type} may be paraphrased as:
\begin{quote}
the time variation of the mean of any observable quantity $\varphi$ (left-hand side) is due, on average, to the mean variation of $\varphi$ in a representative interaction (right-hand side).
\end{quote}
The statement ``on average'' translates the integration with respect to $f(t,\,v)f(t,\,v_\ast)\,dv\,dv_\ast$ on the right-hand side of~\eqref{eq:boltzmann-type}.

Finally, we point out that the coefficient $B>0$ appearing in~\eqref{eq:boltzmann-type} is the \textit{interaction kernel}. It fixes the frequency at which two individuals interact and may be either constant or variable with the opinions $v,\,v_\ast$ of the interacting individuals. In the classical Boltzmann equation~\eqref{eq:boltzmann} the role of a non-constant interaction frequency is played by the collision kernel $B(\bv,\,\bv_\ast):=\abs{(\bv-\bv_\ast)\cdot\bn}$.

In general, it is rather complicated to solve~\eqref{eq:boltzmann-type} or to get qualitative information on its solutions. Nevertheless,~\eqref{eq:boltzmann-type} can be quite easily simulated by means of a computer. Indeed, Algorithm~\ref{alg:nanbu} reported on page~\pageref{alg:nanbu} turns out to be a simple-to-implement and effective method for approximating numerically the solution of~\eqref{eq:boltzmann-type} in any programming language. The reason is that, as shown in Appendix~\ref{sect:derivation}, the derivation of~\eqref{eq:boltzmann-type} follows closely the particle dynamics expressed by Algorithm~\ref{alg:nanbu}. The latter is called \textit{Nanbu-Babovsky Monte Carlo scheme}, see~\cite{bobylev2000PRE,pareschi2001ESAIMP,pareschi2013BOOK} for a more detailed introduction.

The above physical interpretation of~\eqref{eq:boltzmann-type} suggests also a way to generalise the Boltzmann-type equation to the case of \textit{multiple interactions}, i.e. interactions among more than two individuals at a time. Let us consider $M$ interacting individuals and let us denote by $v_i$, $i=1,\,\dots,\,M$, their pre-interaction opinions. The post-interaction opinions will be given by interaction rules of the form
$$ v_i'=v_i+I_i(v_1,\,\dots,\,v_M), \qquad i=1,\,\dots,\,M. $$
The mean variation of an observable quantity in a representative interaction is now
$$ \frac{1}{M}\sum_{i=1}^{M}\left(\varphi(v_i')-\varphi(v_i)\right), $$
therefore, on average, the time variation of the mean of $\varphi$ will obey the equation
$$ \frac{d}{dt}\int_{\sV}\varphi(v)f(t,\,v)\,dv
	=\frac{1}{M}\int_{\sV^M}B\sum_{i=1}^{M}\left(\varphi(v_i')-\varphi(v_i)\right)f(t,\,v_1)\cdots f(t,\,v_M)\,dv_1\,\dots\,dv_M. $$
Here we have used the Boltzmann Ansatz in the form $f_M(t,\,v_1,\,\dots,\,v_M)=f(t,\,v_1)\cdots f(t,\,v_M)$, after observing that in this case we should have used the $M$-joint distribution $f_M$ of the opinions of the interacting individuals. In particular, in the applications that we will discuss in the next sections we will use $M=3$. Then, the multiple-interaction Boltzmann-type equation becomes explicitly
\begin{align}
	\begin{aligned}[b]
		& \frac{d}{dt}\int_{\sV}\varphi(v)f(t,\,v)\,dv \\
		& =\frac{1}{3}\int_{\sV^3}B\left(\varphi(v')+\varphi(v_\ast')+\varphi(v_{\ast\ast}')-\varphi(v)-\varphi(v_\ast)-\varphi(v_{\ast\ast})\right)
			f(t,\,v)f(t,\,v_\ast)f(t,\,v_{\ast\ast})\,dv\,dv_\ast\,dv_{\ast\ast},
	\end{aligned}
	\label{eq:boltz.mult}
\end{align}
where, to restore a notation more in line with the classical one, we have denoted by $v,\,v_\ast,\,v_{\ast\ast}\in\sV$ the pre-interaction opinions of the three interacting individuals.

We refer the interested reader to the recent papers~\cite{toscani2019PRE,toscani2020NHM} for further applications of multiple-interaction Boltzmann-type models to the analysis of various socio-economic problems.

\section{Opinion formation: revisiting the Sznajd model}
\label{sect:sznajd_revisited}
As it often happens in Applied Mathematics, the inspiration for modelling ``new'', viz. non-classical, dynamics, such as e.g. those involving human behaviour, may be borrowed from a mathematical model of a classical, though entirely different, physical system. It is the case of the \textit{Sznajd model} for opinion formation~\cite{sznajd-weron2000IJMP}, that we are going to discuss in this section.

The starting point to approach the Sznajd model is actually the \textit{Ising model}~\cite{ising1925ZFP}, which describes in a stylised but representative way the magnetism of the matter. The Ising model assumes that the atoms of the matter are ordered in a spatial lattice and interact with each other depending on their proximity on this lattice. Each interaction changes one atom's \textit{spin}, namely a discrete variable taking only the two values $\pm 1$. Also the temperature may induce spin changes, however differently from the interactions. The temperature produces indeed random spin fluctuations while the interactions tend to align the spins of the interacting atoms. This simple model allows one to study the transition from the \textit{ferromagnetism}, i.e. when in the long run the interactions prevail collectively on the thermal fluctuations, to the \textit{paramagnetism}, i.e. when the thermal fluctuations dominate.

The analogy with an opinion formation scenario is now clear: the atoms are individuals and the spatial lattice becomes a \textit{social lattice}, for instance a network of contacts. Furthermore, the spin becomes the \textit{opinion}, which in the simplest case may still take the two values $\pm 1$ denoting two opposite choices, for instance yes/no in a referendum. The proximity of the atoms in the spatial lattice becomes a social proximity among the individuals, which allows them to influence each other. Finally, the thermal fluctuations of the spin have their equivalent in the so-called \textit{self-thinking}, i.e. the tendency of the individuals to erratically change opinion independently of the interactions with other individuals~\cite{bennaim2005EL}. However, we anticipate that we will neglect this specific aspect in the following.

In order to build an evolutionary model of opinion formation with these ingredients, it is necessary to specify the elementary rules by which individuals change their opinions when they interact. In this context, these rules will play the same role as the laws~\eqref{eq:collisions}, however with the remarkable difference that they are not suggested by elementary physical properties of the system at hand but are postulated heuristically. This is necessary in view of the new, i.e. non-classical, dynamics we are confronted with, for which, unlike the case of the gas molecules, formal background theories do not exist.

The inspiring principle of the original Sznajd model~\cite{sznajd-weron2000IJMP} may be expressed by the motto ``united we stand'', which summarises the idea that only clusters of similar opinions can spread across the individuals. In particular, following the interpretation given in~\cite{slanina2003EPJB}, we imagine that if two individuals share the same opinion they are able to convince a third individual to change his/her mind and embrace their opinion. This model, which might be called ``two against one'', involves three actors. Denoting their pre-interaction opinions by $v$, $v_\ast$, $v_{\ast\ast}$, the interaction rule just described may be formalised as:
\begin{align}
	\begin{aligned}[c]
		& \text{if } v=v_\ast \text{ then}
			\begin{cases}
				v'=v \\
				v_\ast'=v \\
				v_{\ast\ast}'=v
			\end{cases} \\
		& \text{if } v\ne v_\ast \text{ then no interaction occurs,}
	\end{aligned}
	\label{eq:2.vs.1}
\end{align}
where the statement ``no interaction occurs'' means that each individual maintains his/her pre-interaction opinion unchanged. Actually, we should also take into account that such an opinion exchange takes place on the aforesaid social lattice, thus the individuals $v$, $v_\ast$ need in principle to be neighbours and $v_{\ast\ast}$ needs to be one of their common neighbours on the lattice. Nevertheless, here we will deliberately neglect this aspect for the sake of simplicity.

To pass from the description~\eqref{eq:2.vs.1} of a local representative interaction to that of the aggregate distribution of the opinions, we rely on the kinetic distribution function $f(t,\,v)$ with $v\in\sV=\{-1,\,1\}$. Let $p(t)\in [0,\,1]$ be the percentage of individuals who at time $t$ express the opinion $v=1$ and $q(t)\in [0,\,1]$ that of the individuals who at time $t$ express the opinion $v=-1$. Clearly, $q(t)=1-p(t)$ and furthermore we can represent $f$ as
\begin{equation}
	f(t,\,v)=p(t)\delta(v-1)+q(t)\delta(v+1),
	\label{eq:f.2delta}
\end{equation}
where $\delta(v\pm 1)$ denotes the Dirac delta centred at $\pm 1$, respectively. In other words, since the opinions can take only two values, their probability distribution $f$ is a discrete one and is concentrated only in those two values. Now, the model for the evolution of such an $f$ is provided by the Boltzmann-type equation~\eqref{eq:boltz.mult} with three interacting individuals, where we may plug the expression~\eqref{eq:f.2delta} along with encoding the interaction rules~\eqref{eq:2.vs.1}. Concerning this, note that we need to specify an interaction kernel which vanishes whenever the first two individuals have different opinions, for in such a case the rule~\eqref{eq:2.vs.1} prescribes that there is no interaction as there is no common opinion to spread. Using the characteristic function, we set
\begin{equation}
	B(v,\,v_\ast,\,v_{\ast\ast})=\chi(v=v_\ast)=
	\begin{cases}
		1 & \text{if } v=v_\ast \\
		0 & \text{otherwise}.
	\end{cases}
	\label{eq:B}
\end{equation}
Usually, one says that this transforms~\eqref{eq:boltz.mult} in a Boltzmann-type equation \textit{with cutoff}, because such a $B$ cuts off the cases $v\ne v_\ast$ from the set of the effective interactions.

All of these ingredients together allow us to obtain, in the end, the following system of equations describing the evolution in time of the percentages $p,\,q$:
\begin{equation}
	\begin{cases}
		\dfrac{dp}{dt}=\dfrac{1}{3}pq(p-q) \\[3mm]
		\dfrac{dq}{dt}=\dfrac{1}{3}pq(q-p).
	\end{cases}
	\label{eq:pq}
\end{equation}
Clearly, we need to complement it with an initial condition
$$ p(0)=p_0\in [0,\,1], \qquad q(0)=q_0=1-p_0 $$
which models the initial statistical distribution of the opinions in the society.

\begin{figure}[!t]
\centering
\subfigure[]{\includegraphics[width=0.5\textwidth]{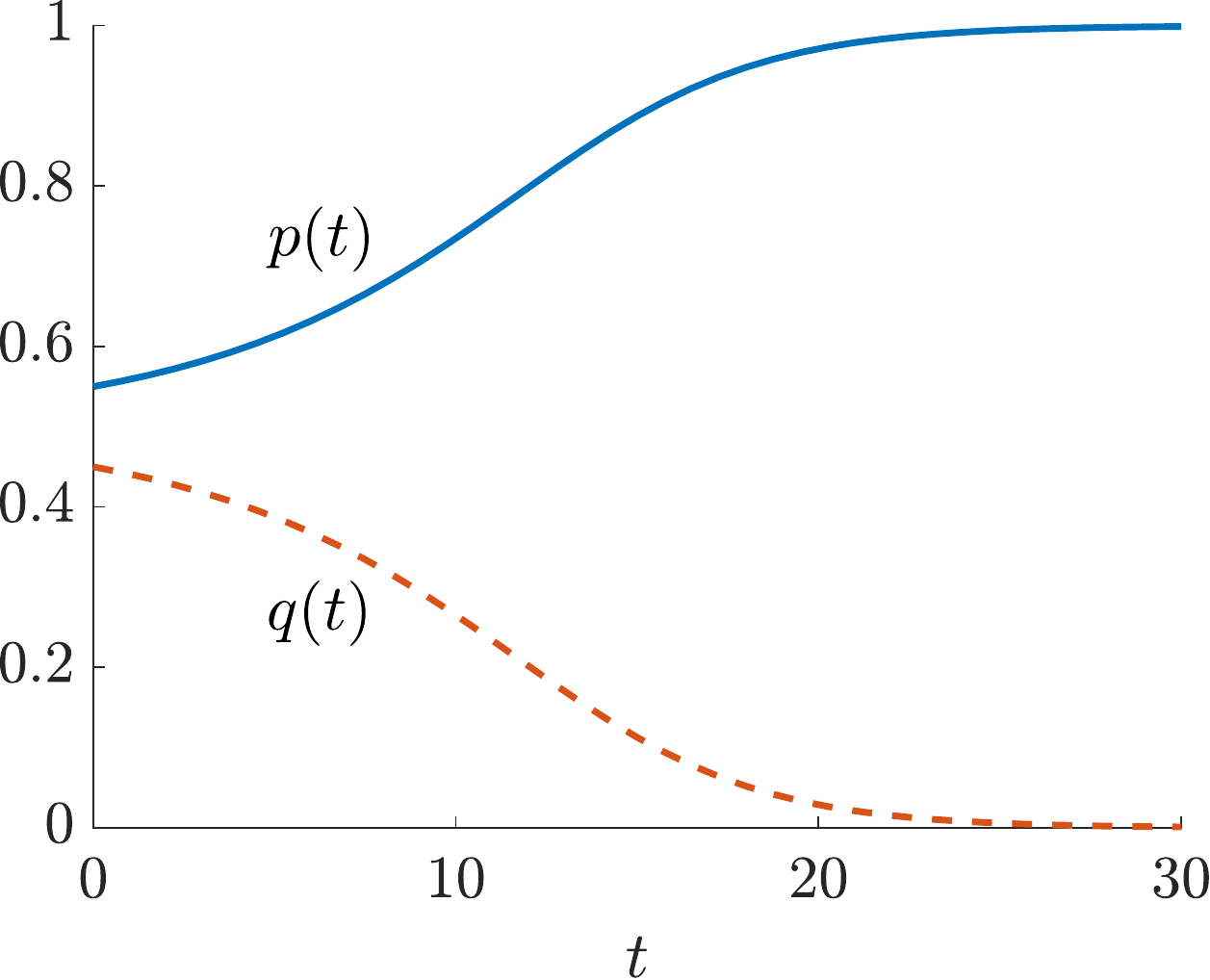}} \\
\subfigure[]{\includegraphics[width=\textwidth]{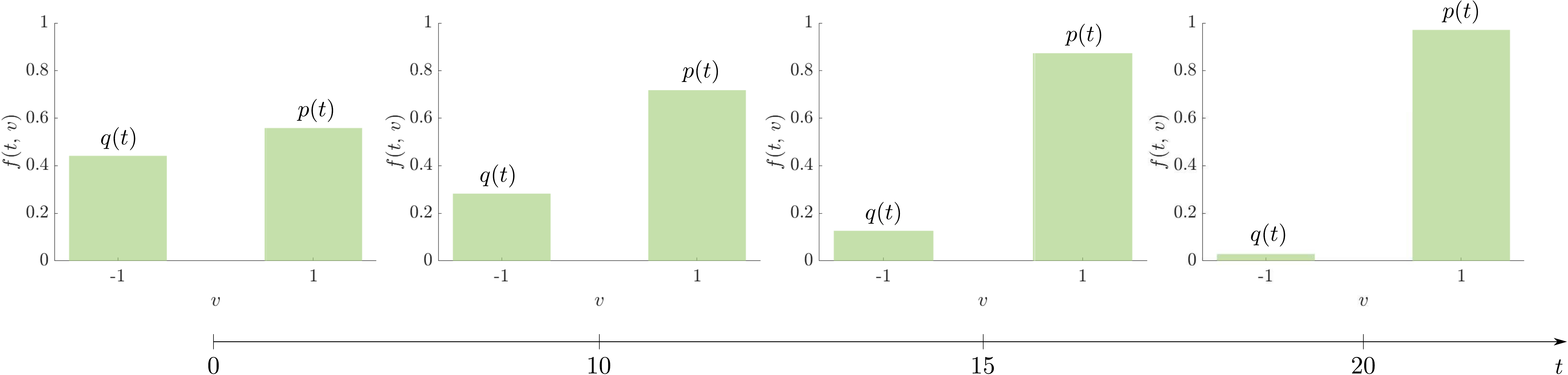}}
\caption{(a) The solution of~\eqref{eq:pq} issuing from $p_0=55\%$ and $q_0=45\%$, thus with an initial predominance of the opinion $v=1$ over $v=-1$. (b) The corresponding statistical distribution of the opinions at different times obtained solving~\eqref{eq:boltz.mult} by means of Algorithm~\ref{alg:nanbu} with $N=3\cdot 10^5$ particles.}
\label{fig:p2}
\end{figure}

From~\eqref{eq:pq}, substituting $q=1-p$, we deduce the following equation in the only unknown $p$:
\begin{equation}
	\frac{dp}{dt}=\frac{1}{3}p(1-p)(2p-1),
	\label{eq:p}
\end{equation}
which makes it easier to infer some stylised facts about the spreading of the opinions in the population. In particular, it is interesting to determine the large time behaviour of the solutions to~\eqref{eq:pq}, which depicts the aggregate trends emerging spontaneously from the elementary interactions among the individuals. In terms of the original Boltzmann-type equation~\eqref{eq:boltz.mult}, this amounts to studying the distribution function to which the unknown $f(t,\,v)$ tends for $t\to +\infty$. To maintain the conceptual parallelism with the classical Boltzmann equation~\eqref{eq:boltzmann} of gas dynamics, it is the counterpart of the so-called \textit{Maxwellian distribution}, namely the distribution of the speeds of the molecules when the gas reaches a statistical equilibrium.

We notice that~\eqref{eq:p} has three equilibria: $p=0$, $p=\frac{1}{2}$, $p=1$, whose the first and third are stable and attractive while the second is unstable\footnote{This can be easily seen by studying the sign of the right-hand side of~\eqref{eq:p}, which gives the sign of the derivative $\frac{dp}{dt}$.}. Unstable equilibria are typically not observed in reality, for they correspond to states that the system cannot maintain: every small perturbation drives spontaneously the system far from those states towards stable ones. The state $p=50\%$ corresponds to a fifty-fifty scenario, in which both opinions $v=\pm 1$ are equally expressed in the society. According to the model, this configuration can be observed in time only if the system starts exactly from $p_0=q_0=50\%$. On the other hand, since this state is unstable, as soon as either $p_0<50\%$ or $p_0>50\%$ the solution to~\eqref{eq:p} is attracted towards the stable states $p=0\%$ or $p=100\%$, respectively, which express instead the tendency of either opinion to predominate in the long run. Figure~\ref{fig:p2} exemplifies these dynamics in the case $p_0=55\%>50\%$, and correspondingly $q_0=45\%<50\%$. In panel (a) we observe the functions $p(t),\,q(t)$ solving~\eqref{eq:pq}, which illustrate the time trends of the percentages of individuals expressing the opinions $v=1$ and $v=-1$, respectively. The solution has been obtained numerically via the fourth-order Runge-Kutta method applied to the ODE~\eqref{eq:p}. In panel (b) we observe instead the kinetic distribution function~\eqref{eq:f.2delta} solving~\eqref{eq:boltz.mult}, which depicts the statistical distribution of the opinions in the society at successive times. This solution has been obtained numerically by means of Algorithm~\ref{alg:nanbu}, sampling initially $N=3\cdot 10^5$ opinions distributed according to the aforesaid percentages $p_0,\,q_0$. By comparing panels (a) and (b), it is evident that the intuitive ``empirical'' procedure at the basis of Algorithm~\ref{alg:nanbu}, which is easily implementable also in complicated cases, is perfectly consistent with the accurate solution of the exact Boltzmann-type equation~\eqref{eq:boltz.mult}, which instead may not be always accessible.

\section{Interlude: Is this model too simple?}
\label{sect:interlude}
One may question that the scenarios depicted by the ``two against one'' dynamics are too poor to represent reliably the opinion trends in a society. Indeed, apart from the unstable configuration $p=q=50\%$, in the long run the model predicts always the emergence of a universal consensus among the individuals on either choice $v=\pm 1$. To make the model more realistic, one might be tempted to complicate the interaction rules~\eqref{eq:2.vs.1}, for instance trying to include more sophisticated behavioural aspects. In general, however, this is not a good idea. Complicated interaction rules often make the model by far less amenable to qualitative investigations, with the result that it may become quickly impossible to understand the fundamental links between the individual and collective behaviour that the model was supposed to unravel.

Many mathematicians struggled to convey the idea that, despite the complexity of reality, simple mathematical models of real world phenomena are usually better than complicated ones. For instance, Martin A. Nowak and Charles R. M. Bangham stated~\cite{nowak1996SCIENCE}:
\begin{quotation}
The strategy of successful mathematical modeling is akin to Ockham's razor: start with the smallest number of essential assumptions and follow the implications rigorously to their logical conclusions. An elegant model can often have greater intrinsic value than an accurate one overloaded with detail. Mathematical models differ from verbal theories in giving a precise and explicit connection between assumption and conclusion.
\end{quotation}

On the same line of thought, George E. P. Box wrote~\cite{box1976JASA}:
\begin{quotation}
Since all models are wrong the scientist cannot obtain a ``correct'' one by excessive elaboration. On the contrary following William of Occam he should seek an economical description of natural phenomena. Just as the ability to devise simple but evocative models is the signature of the great scientist so overelaboration and overparameterization is often the mark of mediocrity.
\end{quotation}
and also elsewhere~\cite{box1979BOOKCH}:
\begin{quotation}
Now it would be very remarkable if any system existing in the real world could be \textit{exactly} represented by any simple model. However, cunningly chosen parsimonious models often do provide remarkably useful approximations. For example, the law $PV=RT$ relating pressure $P$, volume $V$ and temperature $T$ of an ``ideal'' gas via a constant $R$ is not exactly true for any real gas, but it frequently provides a useful approximation and furthermore its structure is informative since it springs from a physical view of the behavior of gas molecules. For such a model there is no need to ask the question ``Is the model true?''. If ``truth'' is to be the ``whole truth'' the answer must be ``No''. The only question of interest is ``Is the model illuminating and useful?''.
\end{quotation}

These considerations apply particularly well to kinetic models, whose ultimate goal is often not (only) to reproduce as much accurately as possible the empirical trajectories of a system. On the contrary, it is first and foremost to explain which stylised types of collective trends emerge spontaneously from elementary individual causes and how the aggregate properties of the former relate to the fundamental characteristics of the latter.

In this respect, the model discussed in Section~\ref{sect:sznajd_revisited} is certainly too simple to predict the great variety of opinion configurations that may arise in reality. Yet it is sufficient to reveal how differently a \textit{majority decision} local rule impacts collectively compared to other conceivable interaction dynamics. For instance, if in place of~\eqref{eq:2.vs.1} we consider the simpler interaction
\begin{equation}
	\begin{cases}
		v'=v \\
		v_\ast'=v,
	\end{cases}
	\label{eq:ochrombel}
\end{equation}
which is inspired by the Ochrombel model~\cite{ochrombel2001IJMP} and describes a situation in which every individual is able to convince whoever else, then from the Boltzmann-type equation~\eqref{eq:boltzmann-type} we get that the evolution of the statistical distribution~\eqref{eq:f.2delta} of the opinions is now such that
$$ \frac{dp}{dt}=\frac{dq}{dt}=0. $$
Hence in this case the opinion distribution does not change in time. In other words, this means that unlike~\eqref{eq:2.vs.1} the interaction rules~\eqref{eq:ochrombel} fail to move the collective opinion, a conclusion that may appear obvious \textit{a posteriori} but which shows that the model is informative.

\section{The effect of indecisiveness/abstention}
\label{sect:indecisiveness}
From the conceptual point of view, it may be useful to enrich the ``two against one'' model not to pursue realistic empirical shapes of the opinion distributions but rather to explore and understand some other common situations not included in the setting of Section~\ref{sect:sznajd_revisited}. For instance, we may want to allow for indecisive people, who cannot make a clear choice between $v=\pm 1$ and prefer therefore to abstain. They may be represented by a third opinion, say $v=0$, and we may denote their percentage in the society by $r=r(t)\in [0,\,1]$. Consequently, the statistical distribution of the opinions becomes now a discrete distribution concentrated in the three values $-1$, $0$, $1$, which we may express as a suitable modification of~\eqref{eq:f.2delta}:
\begin{equation}
	f(t,\,v)=p(t)\delta(v-1)+r(t)\delta(v)+q(t)\delta(v+1),
	\label{eq:f.3delta}
\end{equation}
clearly with $p(t)+r(t)+q(t)=1$ at all times.

Interestingly, no other modifications are required to the modelling setting of Section~\ref{sect:sznajd_revisited} to study this new realistic case. In particular, the interaction rules~\eqref{eq:2.vs.1} as well as the interaction kernel~\eqref{eq:B} remain the same. It is a great advantage when a modelling structure is so sound that with very small changes it can describe several different scenarios. Plugging the expression~\eqref{eq:f.3delta} of $f$ into~\eqref{eq:boltz.mult} we get now the following model of the time evolution of $p$, $r$, $q$:
\begin{equation}
	\begin{cases}
		\dfrac{dp}{dt}=\dfrac{1}{3}p\left[p(r+q)-r^2-q^2\right] \\[3mm]
		\dfrac{dr}{dt}=\dfrac{1}{3}r\left[r(p+q)-p^2-q^2\right] \\[3mm]
		\dfrac{dq}{dt}=\dfrac{1}{3}q\left[q(p+r)-p^2-r^2\right],
	\end{cases}
	\label{eq:pqr}
\end{equation}
which needs to be complemented with an initial condition
$$ p(0)=p_0\in [0,\,1], \qquad r(0)=r_0\in [0,\,1], \qquad q(0)=q_0=1-p_0-r_0 $$
accounting for the initial statistical distribution of the opinions including the indecisive people. Clearly, the values of $p_0,\,r_0$ have to be chosen in such a way that also $q_0$ is comprised between $0$ and $1$, which is obtained for $r_0\leq 1-p_0$.

\begin{figure}[!t]
\centering
\includegraphics[width=0.5\textwidth]{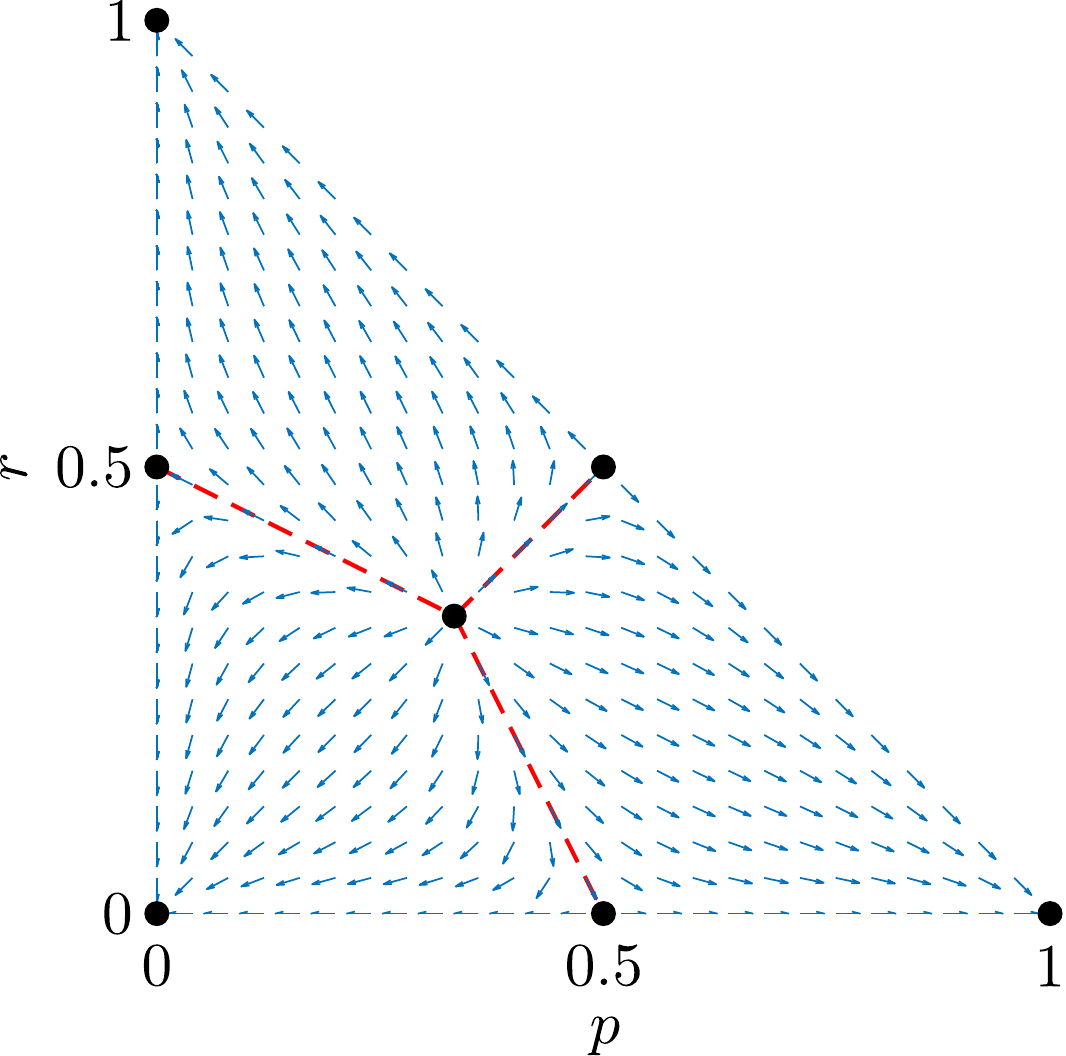}
\caption{The vector field of the differential system~\eqref{eq:pr} and its equilibria.}
\label{fig:vect.field}
\end{figure}

Using $q=1-p-r$, we can drop the third equation in~\eqref{eq:pqr} obtaining the following reduced system in the only unknowns $p$, $r$:
\begin{equation}
	\begin{cases}
		\dfrac{dp}{dt}=\dfrac{1}{3}p\left[p(1-p)-r^2-(1-p-r)^2\right] \\[3mm]
		\dfrac{dr}{dt}=\dfrac{1}{3}r\left[r(1-r)-p^2-(1-p-r)^2\right],
	\end{cases}
	\label{eq:pr}
\end{equation}
which is meaningful for $0\leq p,\,r\leq 1$ and $r\leq 1-p$. These restrictions identify in the plane $p$-$r$ the triangle illustrated in Figure~\ref{fig:vect.field}. This figure shows also the vector field defined by the right-hand side of~\eqref{eq:pr}, which gives the local direction of the trajectories of the system, and its equilibria, i.e. the points where the vector field vanishes. As usual, the equilibria identify the possible states towards which the system evolves in time, hence the possible opinion distributions emerging in the long run from the interactions.

By inspecting the vector field\footnote{Obviously, these results can also be obtained analytically by studying the eigenvalues and eigenvectors of the Jacobian matrix of the vector field of system~\eqref{eq:pr}.} plotted in Figure~\ref{fig:vect.field} we infer that there are three stable equilibria:
$$ (p,\,r,\,q)=(0,\,0,\,1), \qquad (p,\,r,\,q)=(1,\,0,\,0), \qquad (p,\,r,\,q)=(0,\,1,\,0) $$
and four unstable equilibria:
\begin{align*}
	& (p,\,r,\,q)=\left(\frac{1}{3},\,\frac{1}{3},\,\frac{1}{3}\right), \\
	& (p,\,r,\,q)=\left(\frac{1}{2},\,0,\,\frac{1}{2}\right), \qquad
		(p,\,r,\,q)=\left(\frac{1}{2},\,\frac{1}{2},\,0\right), \qquad
			(p,\,r,\,q)=\left(0,\,\frac{1}{2},\,\frac{1}{2}\right).
\end{align*}

\begin{figure}[!t]
\centering
\subfigure[]{\includegraphics[width=0.5\textwidth]{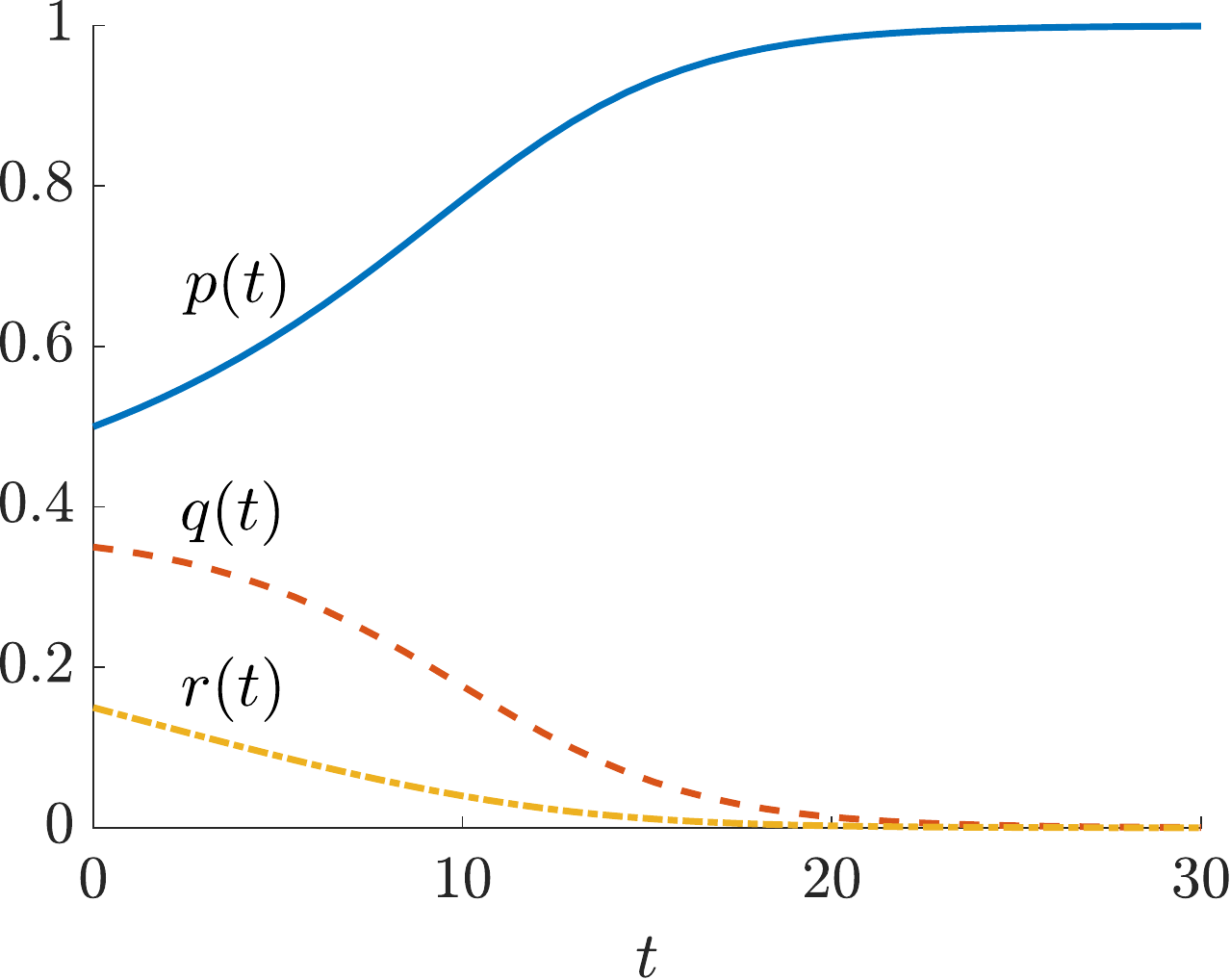}} \\
\subfigure[]{\includegraphics[width=\textwidth]{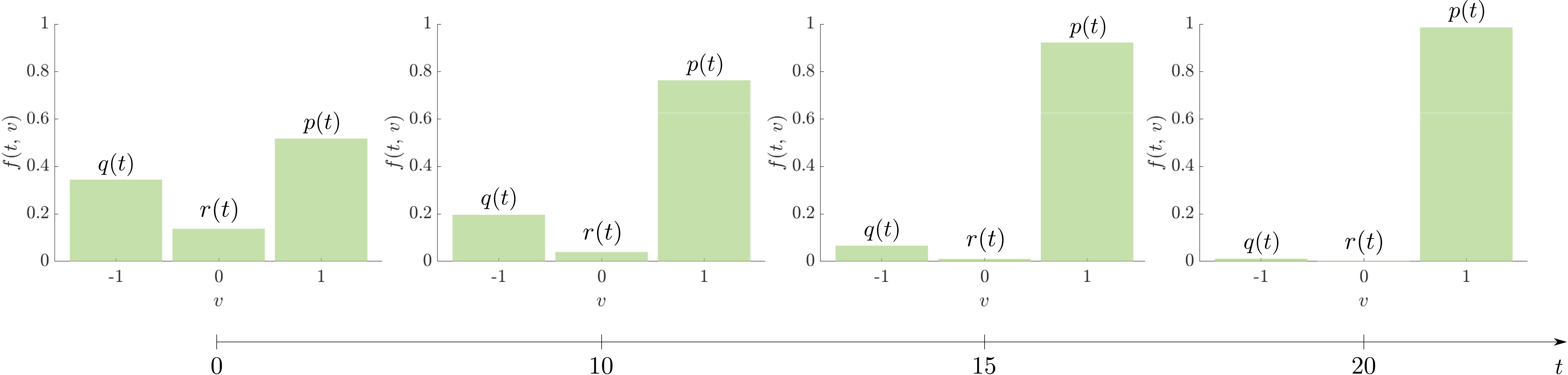}}
\caption{(a) The solution of~\eqref{eq:pqr} issuing from $p_0=50\%$, $q_0=35\%$ and $r_0=15\%$, thus with an initial predominance of the opinion $v=1$ over both $v=-1$ and $v=0$, respectively. (b) The corresponding statistical distribution of the opinions at different times obtained solving~\eqref{eq:boltz.mult} by means of Algorithm~\ref{alg:nanbu} with $N=3\cdot 10^5$ particles.}
\label{fig:p3-1}
\end{figure}

The three stable equilibria correspond to the three situations in which, in the long run, all the individuals agree on one of the three opinions, in particular the one which was initially mostly expressed. For instance, the equilibrium $(p,\,r,\,q)=(1,\,0,\,0)$, which corresponds to the configuration in which the percentage of individual expressing the opinion $v=1$ is $p=100\%$ while the percentages of the individuals expressing the opinions $v=0$ and $v=-1$ are $r=q=0\%$, attracts all the trajectories issuing from the south-east region of the triangle in Figure~\ref{fig:vect.field}, where indeed $p>r,\,q$. An example of this is provided in Figure~\ref{fig:p3-1}. Likewise, the equilibrium $(p,\,r,\,q)=(0,\,0,\,1)$ attracts all the trajectories issuing from the south-west region of the triangle in Figure~\ref{fig:vect.field}, where $q>p,\,r$; and the equilibrium $(p,\,r,\,q)=(0,\,1,\,0)$ attracts all the trajectories issuing from the north region of the triangle, where $r>p,\,q$. These dynamics are not substantially different from those already encountered in the model with only two opinions. In essence, they confirm that with the ``two against one'' interactions the initially dominant opinion tends to attract the whole consensus in the long run.

Also the fact that the uniformly distributed equilibrium $(p,\,r,\,q)=(\frac{1}{3},\,\frac{1}{3},\,\frac{1}{3})$ is unstable is not surprising in view of the case with two opinions: as a matter of fact, it is the counterpart of the configuration $(p,\,q)=(\frac{1}{2},\,\frac{1}{2})$ discussed in Section~\ref{sect:sznajd_revisited}. In the ideal situation in which the three opinions are equally expressed, their statistical distribution is in equilibrium: the interactions shuffle at most the individuals on the various opinions but the percentages are preserved. However, as soon as this configuration is slightly perturbed, e.g. by external factors, so that one of the three opinions dominates such an equilibrium is lost and the system evolves towards one of the asymptotic configurations described above.

\begin{figure}[!t]
\centering
\subfigure[]{\includegraphics[width=0.5\textwidth]{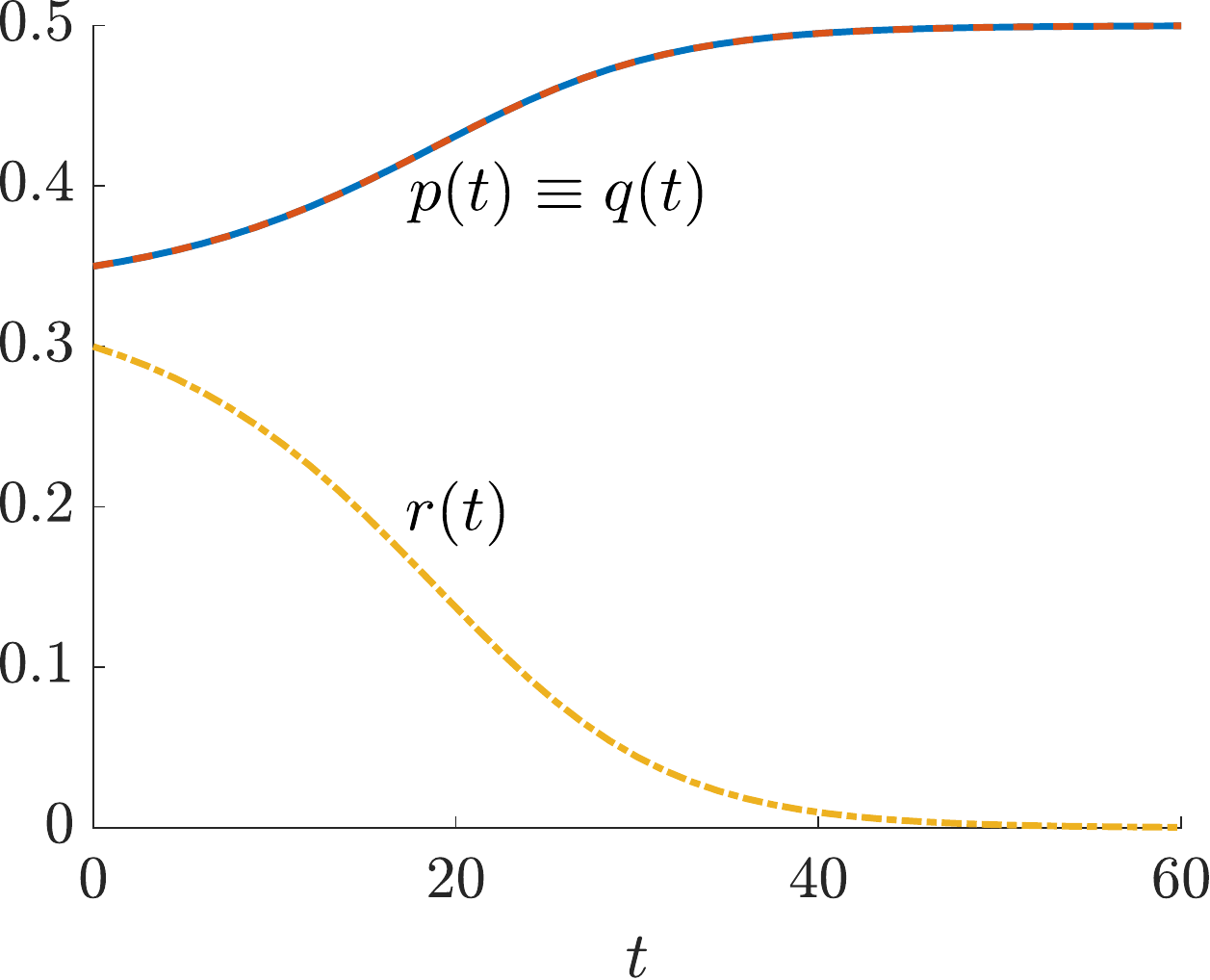}} \\
\subfigure[]{\includegraphics[width=\textwidth]{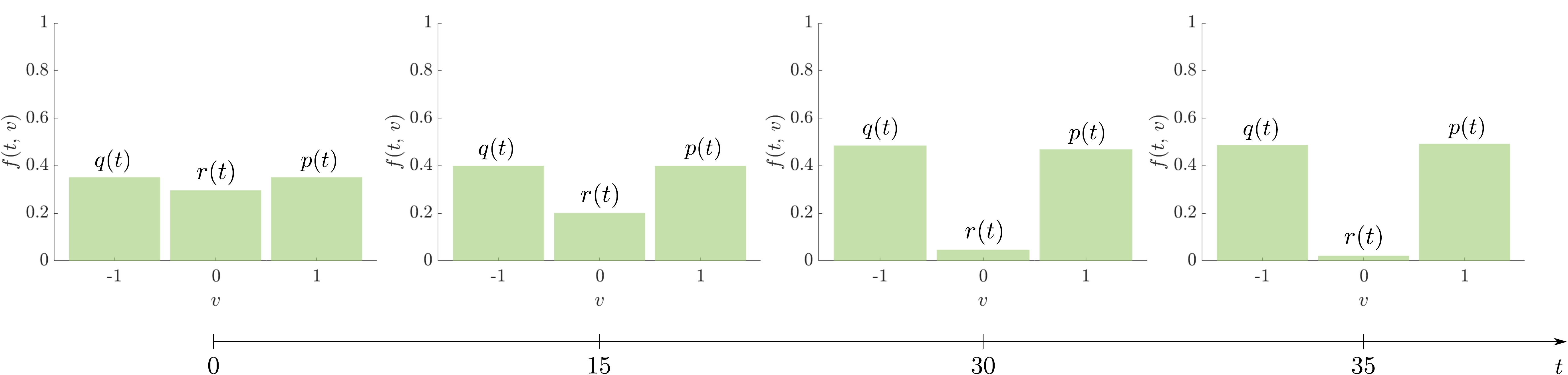}}
\caption{(a) The solution of~\eqref{eq:pqr} issuing from $p_0=q_0=35\%$ and $r_0=30\%$, thus with the same initial predominance of the opinions $v=\pm 1$ over $v=0$. (b) The corresponding statistical distribution of the opinions at different times obtained solving~\eqref{eq:boltz.mult} by means of Algorithm~\ref{alg:nanbu} with $N=12\cdot 10^6$ particles.}
\label{fig:p3-2}
\end{figure}

What makes really the difference with respect to the case with two opinions is the fact that now there may be initially two simultaneously dominant opinions. This is clearly impossible with two opinions whereas with three opinions it may happen, for instance, that at the beginning $v=\pm 1$ are equally expressed in the society while $v=0$ is slightly underexpressed. This corresponds to $p_0=q_0>r_0$, an example of which is illustrated in Figure~\ref{fig:p3-2}. We observe that the opinion $v=0$ disappears in time and the individuals who initially expressed it are progressively convinced to embrace either opinion $v=\pm 1$ in the same proportion. Indeed, the opinions $v=\pm 1$ evolve equally towards a fifty-fifty configuration corresponding to the equilibrium $(p,\,r,\,q)=(\frac{1}{2},\,0,\,\frac{1}{2})$. In the triangle of Figure~\ref{fig:vect.field} this evolution corresponds to the dashed red line separating the south-west and south-east regions. However the equilibrium $(p,\,r,\,q)=(\frac{1}{2},\,0,\,\frac{1}{2})$ is on the whole unstable. Indeed, along the line $r=0$ it reproduces exactly the fifty-fifty configuration of the model with two opinions: any small perturbation drives the system far from it towards either the stable equilibrium with $p=0\%$ or that with $p=100\%$. From Figure~\ref{fig:p3-2}(b) we may appreciate that the basic instability of the equilibrium $(p,\,r,\,q)=(\frac{1}{2},\,0,\,\frac{1}{2})$ makes it challenging to approximate reliably the opinion distribution with Algorithm~\ref{alg:nanbu}. To reach a satisfactory accuracy in this case we need a sample of as many as $N=12\cdot 10^6$ opinions while in the cases of Figures~\ref{fig:p2}(b),~\ref{fig:p3-1}(b) a sample of only $N=3\cdot 10^5$ opinions, i.e. one order of magnitude lower, was sufficient to catch well stable equilibrium distributions.

Totally analogous considerations hold for the other two equilibria of this type, i.e. $(p,\,r,\,q)=(\frac{1}{2},\,\frac{1}{2},\,0)$ and $(p,\,r,\,q)=(0,\,\frac{1}{2},\,\frac{1}{2})$, which are reached when initially $p_0=r_0>q_0$ and $r_0=q_0>p_0$, respectively. The corresponding trajectories of the system are the dashed red lines which, in Figure~\ref{fig:vect.field}, separate the north region from the south-east and the south-west regions, respectively.

\section*{Epilogue}
The kinetic theory provides a powerful and flexible conceptual paradigm to model interacting multi-agent systems and to unravel the links among their properties at different scales: from that of single individuals, where the elementary dynamics take place, to that of the collectivity, where the aggregate effects of the individual interactions are observable. Besides this descriptive level, such a multiscale knowledge is fundamental to act on the system with the aim of modifying its large-scale trends. For this, \textit{bottom-up control approaches}, which are capable of affecting the decisional strategies of a few agents and are then amplified collectively by the interactions, are largely preferable over less feasible \textit{top-down control approaches}, which instead would require to control directly the collectivity at the macroscopic scale. Indeed individual controls may be implemented in practice whereas, in most normal situations, it is virtually impossible to force a large group of agents to behave as a whole in a prescribed manner. It is for instance the case of the driver-assist or autonomous vehicles, namely vehicles with the ability to take automatic decisions, whose use to make the global traffic flow more fluid is already being tested.

Thanks to its intrinsic features, which we have tried to outline in this paper, the kinetic theory may constitute a valid tool to approach these \textit{multiscale automatic decision problems}, inspired by several socio-economic applications, which will presumably play a role in the mathematical research on Artificial Intelligence. Some proposals in this directions are already available. Here, we cite a few examples~\cite{albi2015CMS,albi2014PTRSA,duering2018EPJB,tosin2019MMS,tosin2020MCRF} for the readers interested in this promising and fascinating research line.

\appendix

\section{Insight: Derivation of the Boltzmann-type equation~\texorpdfstring{\eqref{eq:boltzmann-type}}{}}
\label{sect:derivation}
Like in the case of the gas molecules of Section~\ref{sect:legacy}, the idea is to regard the opinion of an individual at time $t>0$ as a random variable $V_t\in\sV$ distributed according to $f(t,\,v)$, i.e. such that
$$ \P(V_t\in A)=\int_Af(t,\,v)\,dv $$
for every (measurable) set $A\subseteq\sV$. In a given time interval $\Delta{t}>0$, two random individuals with opinions $V_t$, $V_{\ast,t}$ may meet and interact. If they do, they update their opinions according to~\eqref{eq:interactions}. Otherwise, they simply maintain their current opinions. To describe this random process, we introduce a random variable $T$ such that $T=1$ if the individuals meet and interact while $T=0$ if they do not. In particular, we may model $T$ as a Bernoulli random variable, which we may further reasonably assume to take the value $1$ with a probability proportional to the duration $\Delta{t}$ of the observation interval. Hence $T\sim\operatorname{Bernoulli}(B\Delta{t})$, meaning
$$ \P(T=1)=B\Delta{t}, \qquad \P(T=0)=1-B\Delta{t}, $$
where $B>0$ is the \textit{interaction kernel} (also called \textit{interaction rate}). Clearly, $\Delta{t}$ has to be chosen in such a way that $\Delta{t}\leq\frac{1}{B}$, so that $B\Delta{t}$ is indeed a probability. We will see in a moment that this is actually not a severe limitation, because we will be interested in instantaneous variations for $\Delta{t}\to 0^+$.

Notice that the random variable $T$ translates the toss of a coin mentioned in line~\ref{state:decide} of Algorithm~\ref{alg:nanbu} to decide whether the two randomly chosen individuals interact. The coin is biased whenever $B\Delta{t}\ne\frac{1}{2}$.

At this point, we are in a position to write the random encounter-interaction dynamics as:
\begin{subequations}
\begin{align}
	& V_{t+\Delta{t}}=(1-T)V_t+TV'_t \label{eq:Vt} \\
	& V_{\ast,t+\Delta{t}}=(1-T)V_{\ast,t}+TV'_{\ast,t}, \label{eq:V*t}
\end{align}
\end{subequations}
where, in view of~\eqref{eq:interactions}, $V'_t:=V_t+I(V_t,\,V_{\ast,t})$ and $V'_{\ast,t}:=V_{\ast,t}+I_\ast(V_{\ast,t},\,V_t)$. The relationships~\eqref{eq:Vt},~\eqref{eq:V*t} simply mean that, after a time $\Delta{t}$, the opinions of the individuals may or may not have changed depending on whether an interaction actually took place during the time $\Delta{t}$. Let us focus in particular on~\eqref{eq:Vt}. If we take any quantity $\varphi$ which can be computed out of the knowledge of an opinion $v$, i.e. $\varphi=\varphi(v)$, then we clearly have
\begin{equation}
	\varphi(V_{t+\Delta{t}})=\varphi\bigl((1-T)V_t+TV'_t\bigr),
	\label{eq:varphi}
\end{equation}
which trivially generalises the relationship~\eqref{eq:Vt} by saying that, after a time $\Delta{t}$, the value of $\varphi$ may or may not have changed depending on whether an interaction took place during the time $\Delta{t}$. The function $\varphi$ is generally called an \textit{observable quantity}. Let us now compute the average of both sides of~\eqref{eq:varphi} with respect to all the sources of randomness, i.e. $V_t$, $V_{\ast,t}$, $T$. Denoting by $\ave{\cdot}$ such an average and computing explicitly the expectation of the right-hand side with respect to $T$ we discover:
\begin{align*}
	\ave{\varphi(V_{t+\Delta{t}})} &= \ave{\varphi\bigl((1-T)V_t+TV'_t\bigr)} \\
	&= \ave{(1-B\Delta{t})\varphi(V_t)}+\ave{B\Delta{t}\varphi(V'_t)},
\end{align*}
which, rearranging the terms and dividing by $\Delta{t}$, becomes
$$ \frac{\ave{\varphi(V_{t+\Delta{t}})}-\ave{\varphi(V_t)}}{\Delta{t}}=\ave{B(\varphi(V'_t)-\varphi(V_t))}. $$
In the limit $\Delta{t}\to 0^+$, this yields formally
\begin{equation}
	\frac{d}{dt}\ave{\varphi(V_t)}=\ave{B(\varphi(V'_t)-\varphi(V_t))}.
	\label{eq:varphi(Vt)}
\end{equation}
Likewise, repeating the same procedure on~\eqref{eq:V*t} we get
\begin{equation}
	\frac{d}{dt}\ave{\varphi(V_{\ast,t})}=\ave{B(\varphi(V'_{\ast,t})-\varphi(V_{\ast,t}))}
	\label{eq:varphi(V*t)}
\end{equation}
and finally, summing~\eqref{eq:varphi(Vt)} and~\eqref{eq:varphi(V*t)},
\begin{equation}
	\frac{d}{dt}\bigl(\ave{\varphi(V_t)}+\ave{\varphi(V_{\ast,t})}\bigr)=\ave{B\bigl(\varphi(V'_t)+\varphi(V'_{\ast,t})-\varphi(V_t)-\varphi(V_{\ast,t})\bigr)}.
	\label{eq:varphi(Vt)+varphi(V*t)}
\end{equation}

From here, it is now straightforward to deduce an equation for the distribution function $f$. Indeed, considering that $V_t$ and $V_{\ast,t}$ are distributed according to $f(t,\,v)$ (by definition of $f$ itself), we have:
$$ \ave{\varphi(V_t)}=\ave{\varphi(V_{\ast,t})}=\int_{\sV}\varphi(v)f(t,\,v)\,dv $$
while
\begin{multline*}
	\ave{B\bigl(\varphi(V'_t)+\varphi(V'_{\ast,t})-\varphi(V_t)-\varphi(V_{\ast,t})\bigr)} \\
		=\int_{\sV}\int_{\sV}B\bigl(\varphi(v')+\varphi(v_\ast')-\varphi(v)-\varphi(v_\ast)\bigr)f(t,\,v)f(t,\,v_\ast)\,dv\,dv_\ast.
\end{multline*}
In the last equation, $v',\,v_\ast'$ have to be thought of as functions of $v,\,v_\ast$ through~\eqref{eq:interactions}. Moreover, we have used the Boltzmann Ansatz $f_2(t,\,v,\,v_\ast)=f(t,\,v)f(t,\,v_\ast)$. Notice indeed that, in principle, the expectation of the right-hand side of~\eqref{eq:varphi(Vt)+varphi(V*t)} should be computed using the joint distribution $f_2$, because $\varphi(v'),\,\varphi(v_\ast')$ depend jointly on $v,\,v_\ast$. Putting all the elements together, we finally arrive at
\begin{equation}
	\frac{d}{dt}\int_{\sV}\varphi(v)f(t,\,v)\,dv
		=\frac{1}{2}\int_{\sV}\int_{\sV}B\bigl(\varphi(v')+\varphi(v_\ast')-\varphi(v)-\varphi(v_\ast)\bigr)f(t,\,v)f(t,\,v_\ast)\,dv\,dv_\ast.
	\label{eq:boltz-type.weak.asymm}
\end{equation}
This is the \textit{weak form}~\eqref{eq:boltzmann-type} of the equation for $f$, as it has to hold for all possible choices of the observable quantity $\varphi$. The latter plays, in this context, the role of a test function.

A few remarks on~\eqref{eq:boltz-type.weak.asymm} are now in order.
\begin{enumerate}[label=(\roman*)]
\item If the interaction kernel $B$ is constant, it may be clearly written as a coefficient in front of the integrals on the right-hand side. In general, however, $B$ may depend on the states $v,\,v_\ast$ of the interacting individuals, i.e. $B=B(v,\,v_\ast)$.
\item Equation~\eqref{eq:boltz-type.weak.asymm} simplifies as
\begin{equation}
	\frac{d}{dt}\int_{\sV}\varphi(v)f(t,\,v)\,dv
		=\int_{\sV}\int_{\sV}B\bigl(\varphi(v')-\varphi(v)\bigr)f(t,\,v)f(t,\,v_\ast)\,dv\,dv_\ast
	\label{eq:boltz-type.weak.symm}
\end{equation}
if $I=I_\ast$ in~\eqref{eq:interactions} and if moreover $B$ is symmetric, i.e. $B(v,\,v_\ast)=B(v_\ast,\,v)$. Notice that these conditions are satisfied by the collisions~\eqref{eq:collisions}, where $I(\bv,\,\bv_\ast)=((\bv_\ast-\bv)\cdot\bn)\bn$ and $I_\ast(\bv_\ast,\,\bv)=((\bv-\bv_\ast)\cdot\bn)\bn$, and by the collision kernel of the Boltzmann equation~\eqref{eq:boltzmann}, namely $B(\bv,\,\bv_\ast)=\abs{(\bv-\bv_\ast)\cdot\bn}$. From~\eqref{eq:boltz-type.weak.symm}, by the change of variables~\eqref{eq:interactions} and the arbitrariness of $\varphi$, it is possible to deduce the \textit{strong form} of the equation for $f$, which reads
$$ \partial_tf(t,\,v')=\int_{\sV}\left(B(v,\,v_\ast)\frac{1}{J}f(t,\,v)f(t,\,v_\ast)-B(v',\,v_\ast')f(t,\,v')f(t,\,v_\ast')\right)dv_\ast', $$
where now $v,\,v_\ast$ have to be thought of as functions of $v',\,v_\ast'$. Here, $J$ stands for the modulus of the determinant of the Jacobian matrix of the change of variables~\eqref{eq:interactions}. In this form, the analogy with the Boltzmann equation~\eqref{eq:boltzmann} is evident, considering that from the collision rule~\eqref{eq:collisions} it results $J=1$ and $B(\bv,\,\bv_\ast)=B(\bv',\,\bv_\ast')$.
\end{enumerate}

\bibliographystyle{plain}
\bibliography{FmTa-boltz_revisited}

\bigskip\bigskip

\begin{tabular}{m{2cm}m{11cm}}
\includegraphics[width=2cm]{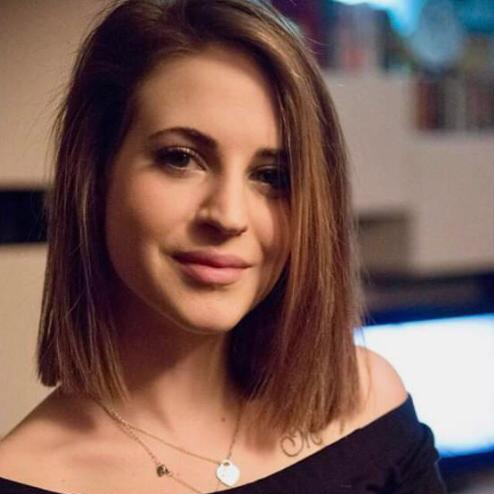} & {\small\textbf{Martina Fraia} graduated in Mathematics for Engineering (``Matematica per l'Ingegneria'') at Politecnico di Torino in March 2020. This paper, in particular the revisitation of the Sznajd model from a Boltzmann-type kinetic perspective and the extension to the case of indecisive people (cf. Section~\ref{sect:indecisiveness}), originates from the contents of her BSc thesis.} \\
\\[5mm]
\includegraphics[width=2cm]{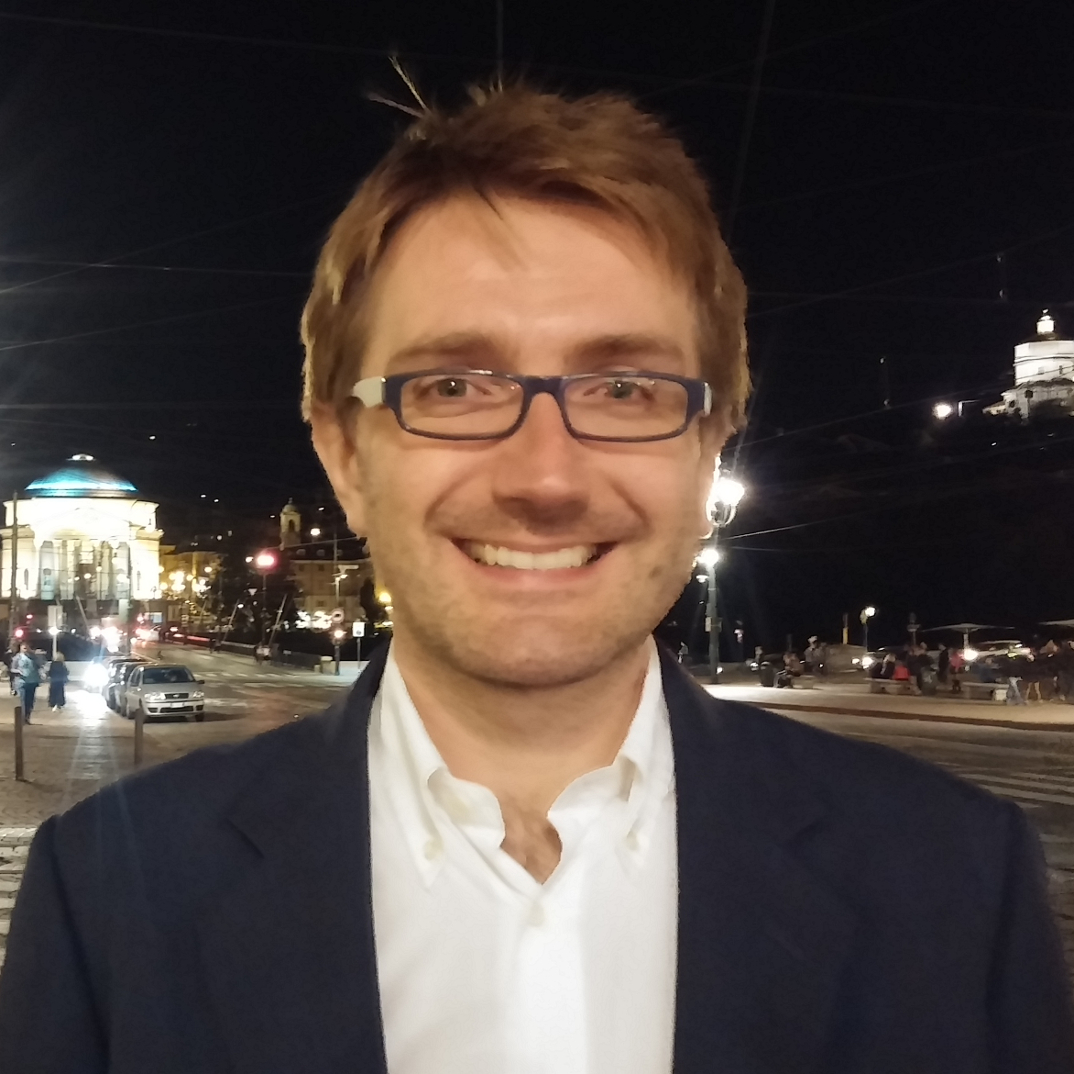} & {\small\textbf{Andrea Tosin} is professor of Mathematical Physics at the Department of Mathematical Sciences ``G. L. Lagrange'' of Politecnico di Torino. His research consists in revisiting the classical methods of kinetic theory to investigate emerging problems in the realm of interacting multi-agent systems, in particular vehicular traffic and social systems.}
\end{tabular}

\end{document}